\documentstyle[12pt,WSPRL,epsfig]{article} 
\begin{document}

\vspace*{1.0cm}

IFP-727-UNC, September 1996

\vspace*{3.0cm}

\centerline{\normalsize\bf Lecture on a $Z^{'}$ Coupled to Beauty and Truth} 
\baselineskip=16pt 
\centerline{\normalsize Paul H. Frampton} 
\centerline{\normalsize Department of Physics and Astronomy,University of North Carolina,}
\centerline{\normalsize  Chapel Hill, NC  27599-3255, U.S.A.} 
\baselineskip=13pt 
\vspace*{10.0cm}

{\it Lecture at Ettore Majorana International School of Subnuclear Physics,}
{\it Erice, Sicily, July 1996.}

\newpage

\centerline{\normalsize\bf Lecture on a $Z^{'}$ Coupled to Beauty and Truth} 
\baselineskip=16pt 
\centerline{\normalsize Paul H. Frampton} 
\centerline{\normalsize Department of Physics and Astronomy,University of North Carolina,}
\centerline{\normalsize  Chapel Hill, NC  27599-3255, U.S.A.} 
\baselineskip=13pt 
\vspace*{0.9cm} 

\begin{abstract} 

By extending the standard gauge group to ${\rm SU}(3)_c \times 
{\rm SU}(2)_L \times {\rm U}(1)_Y \times {\rm U}(1)_X$ 
with $X$ charges carried only by the third family 
we accommodate the LEP measurement of $R_b$ and predict a 
potentially measurable discrepancy in $A_{FB}^{b}$ in $e^+e^-$ 
scattering and that $D^0 \bar D^0$ mixing may be near its 
experimental limit. The uniqueness of our model is that the  
$Z^{'}$ couplings are generation-dependent and hence explicitly 
violate the GIM mechanism, but can nevertheless be naturally 
consistent with FCNC constraints.  Direct detection of this $Z^{'}$ is 
possible but challenging.  

\end{abstract} 

\vspace*{0.9cm} 

Not long after this talk was given in Erice on July 10, 1996, in a plenary session 
at the International Conference on High Energy Physics  on July 30, 1996
in Warsaw, A. Blondel announced significantly-changed experimental data.
In this write-up (September), I follow the original talk but mention 
{\it in italics} the new data.    

\vspace*{0.9cm}

{\bf I. Background and Motivation.}

Although the Standard Model (SM) survived the high 
precision LEP measurements almost unscathed, there are a few 
discrepancies which persist, most of them at a low level of 
statistical significance and hence quite likely to disappear 
as more data are collected.  One outstanding deviation from the 
SM which is quite large involves the couplings of the beauty ($b$) quark.

That it occurs in the third quark family makes it more plausible
because the heaviness of the top quark makes this sector the most suspect.

In particular, the ratio
$R_b = \Gamma(Z \rightarrow b\bar{b})/\Gamma(Z \rightarrow {\rm hadrons})$
is predicted by the SM to be $R_b = 0.2156 \pm 0.0003$~\cite{pdglang} (where
the uncertainty comes from $m_t$ and $m_H$)  and is measured to be
$R_b = 0.2219 \pm 0.0017$~\cite{PDG}, about $3\%$ too high and a significant
$3.7\sigma$ effect (for a recent analysis see Ref.~\cite{BBM}). In this talk,
we shall thus take the $R_b$ data at face value
and construct an extension of the standard model that explains $R_b$ and has
other with testable predictions. The two simplest ways to extend the SM while
preserving its principal
features are to extend the gauge sector or to extend the fermion sector.
In the former approach, the simplest possibility is to extend the
gauge sector by a ${\rm U}(1)$ gauge field which mixes with the usual $Z$
boson and generates non-standard couplings to $b$ quarks and perhaps the
other quarks and leptons. Such an approach was first discussed in
Ref.~\cite{holdom} and in a different context in Ref.~\cite{CR}.
More recently, attempts have been made to explain the $R_b$
and $R_c$ discrepancies with an extra ${\rm U}(1)$ gauge field
which couples also to light quarks~\cite{CA}.
The simplest fermion-mixing model
to explain the $R_b$ (and $R_c$) data was proposed in Ref.~\cite{ma}.

\vspace*{0.7cm}

{\it The new data on $R_b$ are as follows:

\begin{itemize}

\item{ } $R_b$ = 0.2158 $\pm$ 0.0010 $\pm$ 0.0011  (ALEPH)
\item{ } $R_b$ = 0.2178 $\pm$ 0.0028 $\pm$ 0.0027  (DELPHI)
\item{ } $R_b$ = 0.2185 $\pm$ 0.0028 $\pm$ 0.0032  (L3)
\item{ } $R_b$ = 0.2149 $\pm$ 0.0032 $\pm$ 0.0021  (SLD)
\item{ } $R_b$ = 0.2178 $\pm$ 0.0011 (WORLD AVERAGE which is $+1.9\sigma$)

\end{itemize}

}

\vspace*{0.7cm}

{\it Although the discrepancy in $R_b$ lessened, the left-right asymmetry
$A^b$ increased its discrepancy from the SM value:

\begin{itemize}

\item{ } $A^b$ = 0.890 $\pm$ 0.029 ($A_{FB}$ from LEP)
\item{ } $A^b$ = 0.863 $\pm$ 0.049 ($A_{LR}$ from SLD)
\item{ } $A^b$ = 0.867 $\pm$ 0.022 (WORLD AVERAGE which is $-3.0\sigma$)

\end{itemize}

}

\vspace*{0.7cm}

The list of departures from the Standard Model is presently very short:

\begin{itemize}

\item{1. } The question of $R_b$ we are discussing.

\item{2. } Neutrino masses. The problem with $\nu_{\odot}$ is compelling.
There is also the problem with atmospheric neutrinos,  the LSND data, and
the possibility of $\nu_{\tau}$ as a HDM candidate.

\item{3. } Large $E_T$ jets at Fermilab but new data and theory seem to
be resolving much of the problem. 

\item{4. } One event, $e^+e^-\gamma\gamma + \rm{missing} E_T$,  at Fermilab. 
It is very interesting, but seems premature, to draw conclusions on the basis 
of only one event.

\end{itemize}

It is not difficult to find models in which the radiative corrections
can accommodate $R_b$ measurements~\cite{LN,BBCLN}; however, many popular
models fail to provide a convenient solution.
The Minimal Supersymmetric Standard Model
(MSSM) is a notable example of this. Only a small region of parameter space
can yield a consistent result, corresponding to a light supersymmetric
spectrum, detectable at LEP II~\cite{WK,ELN}
(see however Ref.~\cite{FPT} for a light gluino alternative).
Two-Higgs doublet models also fall into this category\cite{LN,TH}.
For a comprehensive review of the possibilities see Ref.~\cite{BBCLN}
and references therein.  We extend the gauge sector
by adopting the choice of gauge group
${\rm SU}(3) \times {\rm SU}(2)_L \times {\rm U}(1)_Y \times {\rm U}(1)_X$.
Associated with the additional
${\rm U}(1)_X$ gauge group is a new quantum number $X$
which defines the strength of the beauty and top couplings to the
one new gauge boson which will be denoted by $Z^{'}$ for simplicity,
although this $Z^{'}$ will certainly couple differently than
any other $Z^{'}$ in the literature. What differentiates our model from
others\cite{holdom,CR,CA,barger} is that the
$Z^{'}$ couplings are generation-dependent, the GIM mechanism is
explicitly broken and yet the FCNC constraints can be naturally
satisfied.

To proceed with presenting our model we
shall first examine the decay
of the $Z$ and its relation to the fundamental
$Z$-fermion couplings of the effective Lagrangian.
The decay of the $Z$ into a fermion-antifermion pair $f\bar{f}$ is given by:

\begin{equation}
\Gamma(Z \rightarrow f\bar{f}) =  \left( \frac{\alpha_{em}(M_Z)C M_Z}
{6c_W^2s_W^2} \right) \beta \Bigl( (g_L^{f2} + g_R^{f2})(1 - x)
+ 6xg_L^fg_R^f \Bigr)~,\label{Zwidth}
\end{equation}
where $c_W = \cos\theta_W$, $g_L^f = T_3^f - Q^f\sin^2\theta_W$,
$g_R^f = -Q^f\sin^2\theta_W$, $x = (m_f/M_Z)^2$ and
$\beta = \sqrt{1 - 4x}$. The color factor is $C = 3$
for quarks and $C = 1$ for leptons. For the light fermions, it is an
adequate approximation to put $x = 0$ and $\beta = 1$ and,
using $\sin^2\theta_W = 0.232$, this gives the familiar values
$\Gamma_e = \Gamma_{\mu} = \Gamma_{\tau} \simeq 83 {\rm\ MeV}$ and
$\Gamma_{\nu_{i}} \simeq 166 {\rm\ MeV}$ for $i = e, \mu, \tau$
and for the quarks, $\Gamma_u = \Gamma_c \simeq 285 {\rm\ MeV}$
and $\Gamma_d = \Gamma_s = \Gamma_b \simeq 367 {\rm\ MeV}$.

The couplings $g^f_{L,R}$ are modified when the $Z$ mixes with a
$Z'$.
The effective Lagrangian for the $Z$ and $Z'$ coupling to fermions is

\begin{equation}
{\cal L}_{eff} =  g_Z Z^{\mu}\bar{f}\gamma_{\mu}(g_L^fP_L + g_R^fP_R)f
+g_XZ^{'\mu}\bar{f}\gamma_{\mu}(X_L^fP_L + X_R^fP_R)f~, \label{Leff}
\end{equation}
where $g_Z = g_2/c_W = 0.739$, and $P_{R,L} =  (1 \pm \gamma_5$)/2. This
$Z^{'}$ does not mix with the photon and
the electric charge still given by $Q = T_3 + Y/2$,
where $Y$ is the hypercharge and $T_3$ the third component of
weak isospin.
The mass eigenstates are mixtures of these states with
a mixing angle according to $\hat{Z} = Z\cos\alpha - Z^{'}\sin\alpha$ and
$\hat{Z}^{'} = Z^{'}\cos\alpha + Z\sin\alpha$. If the mass matrix is given by

\begin{equation}
\left( Z\  Z^{'} \right) \left( \begin{array}{cc} M^2 &  \delta M^{2} \\
\delta M^{2} & M^{'2} \end{array} \right)
\left( \begin{array}{c} Z \\ Z^{'} \end{array} \right),
\end{equation}
then the mixing angle is given by

\begin{equation}
\tan\alpha = \frac{ \delta M^2}{\hat{M}^2_{Z^{'}} - M^2} =
\frac{ \delta M^2}{M^{'2} - \hat{M}^2_Z}~,\label{tana}
\end{equation}
where the hats denote mass eigenvalues.
Because of the level of agreement between the SM and leptonic $Z$ decays
at LEP,   $\cos^2\alpha$ must be near unity.
In fact we find $\rm{cos}^2\alpha > 0.995, \rm{tan}\alpha < 0.07$.

\vspace{0.7cm}

{\bf II. The Model.\cite{wise}}

We extend the gauge group to $SU(3)_c \times SU(2)_L \times U(1)_Y \times U(1)_X$.   
The leptons and quarks u, d, c, s have $X = 0$ as does the standard Higgs
doublet $\phi$. Generally $X_L^b, X_R^b, X_L^t, X_R^t$ may be non-zero, and we add extra Higgs
doublet(s) $\phi^{'}, \phi^{''},....$ for which $X_{\phi^{'}}, X_{\phi^{''}},....$
are non-zero.

In the presence of the $Z^{'}$,
we see from Eq.~(\ref{Leff}) that
the $Z$ couplings are modified according to:

\begin{equation}
\delta g_L^f =  -\frac{g_X}{g_Z} X_L^f \tan\alpha~,\qquad
\delta g_R^f =  -\frac{g_X}{g_Z} X_R^f \tan\alpha~,
\end{equation}
where we have factored out a $\cos\alpha$ factor common
to all the mass eigenstate $\hat{Z}$ couplings.
The change $\delta R_b$ is given at lowest order in the mixing by

\begin{equation}
\delta R_b  =  R_b - R_b^{(0)}
 =  2R_b^{(0)} (1 - R_b^{(0)}) \left( \frac{g_L^{b(0)}
\delta g_L^b + g_R^{b(0)} \delta g_R^b}{(g_L^{b(0)})^2 + (g_R^{b(0)})^2}
\right)~,  \label{Rb}
\end{equation}
where the superscript $0$ denotes SM quantities and
$g_L^{b(0)} = -0.423$ and $g_R^{b(0)} = 0.077$.
Requiring $R_b$ to be within one standard deviation
of the experimental value means that $ 0.0080 > \delta R_b > 0.0046$.
Depending on the ${\rm U}(1)$ charges of the $t$ and $b$ quarks
we consider adding a second ($\phi^{'}$, $X_{\phi^{'}}=+1$) and possibly third
($\phi^{''}, X_{\phi^{''}}= -1$) Higgs doublet
to the SM doublet ($\phi$, $X_{\phi} = 0$).
First consider the case of only {\it two} Higgs doublets.
Here $\phi^{'}$ couples to both $b$ and $t$ and so $X_{\phi^{'}} =
X_L^b - X_R^b = - X_L^t + X_R^t$. Then we can write
$\delta M^2 = - X_{\phi^{'}} g_Xg_Z|\langle \phi^{'} \rangle|^2$
and using Eq.~(\ref{tana})
we see that $X_{\phi^{'}}\tan\alpha < 0$\footnote{We are here
assuming that $\hat{M}_{Z^{'}} > \hat{M}_Z$.
Models with $\hat{M}_Z > \hat{M}_{Z^{'}}$ can be constructed but
their parameter space is more restricted.}. If only $b_L$ or $b_R$
has nonzero $X$ charge then $X_{\phi^{'}} = X_L^b$ or $X_{\phi^{'}}
= - X_R^b$ respectively
and because of the signs of $g_L^{b(0)}$ and $g_R^{b(0)}$ in Eq.~(\ref{Rb}),
$R_b$ would always be {\it decreased}. We
must therefore consider both $X^b_{L,R}$ nonzero. Then we can write (\ref{Rb})
numerically as

\begin{equation}
\delta R_b =  g_X\tan\alpha (1.05\, X_{\phi^{'}} + 0.86\, X^b_R)~,\label{numRb}
\end{equation}
so $-X^b_R/X_{\phi^{'}} >
 1.2$ in order to get a positive effect.  To see
that this is inconsistent, we must
use another constraint: the measured $Z$-pole forward-backward asymmetry
in $e^+e^-\rightarrow \bar{b} b$, $A^{(0,b)}_{FB}$. To leading order
it is given by

\begin{equation}
\delta A_{FB}^{(0,b)} = A_{FB}^{(0,b)} - A_{FB}^{(0,b)(SM)}
= A_{FB}^{(0,b)(SM)}  \frac{4(g_L^{b(0)})^2(g_R^{b(0)})^2}
{(g_L^{b(0)})^4 - (g_R^{b(0)})^4} \left(
\frac{\delta g_L^b}{g_L^{b(0)}}
- \frac{\delta g_R^b}{g_R^{b(0)}} \right)~.\label{Afb}
\end{equation}

Inserting the numerical values, including $A_{FB}^{(0,b)(SM)} = 0.101$,
we find that

\begin{equation}
\delta A_{FB}^{(0,b)} = g_X \tan\alpha (0.043\, X_{\phi^{'}} + 0.278\, X_R^b)~.
\label{numAfb}
\end{equation}
Comparison of the experimental forward-backward asymmetry with the
SM prediction allows only
a small departure satisfying $|\delta A_{FB}^{(0,b)}| < 0.003$~\cite{PDG}.
Using the lowest consistent value of $\delta R_b$ then shows that
$A_{FB}^{(0,b)}$ is too big.
This excludes all models with only the two scalar doublets $\phi$ and
$\phi^{'}$.

So we must add a third doublet $\phi^{''}$ which gives mass
to the $t$ quark, $\phi^{'}$ still coupling to the $b$ quark. Thus
$X_{\phi^{''}} = - X^t_L + X^t_R$ and $X_{\phi^{'}} = X^b_L - X^b_R$.
In this case we have
$\delta M^2 = -g_Xg_Z (X_{\phi^{'}}|\langle \phi^{'} \rangle|^2
+ X_{\phi^{''}}|\langle \phi^{''} \rangle|^2)$ and with opposite signs
for $X_{\phi^{'}}$ and $X_{\phi^{''}}$
and the natural choice $|\langle \phi^{''} \rangle| > |\langle \phi^{'}
\rangle|$
we can make $X_{\phi^{'}}\tan\alpha > 0$.
We are thus free to make simple choices for the quark charges.
There are two natural choices to consider: (i) $X_L^b = 1; X_R^b = 0$
and (ii) $X_L^b = 0; X_R^b = 1$.
Of these, (ii) can be shown to be inconsistent with the data, as follows.
Equations~(\ref{numRb}) and (\ref{numAfb}) give
$\delta R_b = - 0.19\, g_X \tan\alpha $ and
$\delta A_{FB}^{(0,b)} = 0.24\, g_X \tan\alpha$.
Requiring $\delta R_b > 0.0046$, implies
$|\delta A_{FB}^{(0,b)}| > 0.005$ contradicting experiment.
This then leaves our preferred model: the charges for the third family
- defined more carefully below - are simply $X_L^{b,t}=1$ and $X_R^{b,t}=0$.
The model has three Higgs scalar doublets $\phi, \phi^{'}$ and $\phi^{''}$
with $X$ charges $0$, $+1$ and $-1$ respectively. (There will also be
Higgs scalars with X charge but no standard model quantum numbers).

Cancellation of chiral anomalies is most economically accomplished by
adding two doublets of quarks $(w, w^{'})_L + (w, w^{'})_R$ which are
vector-like in weak hypercharge. The doublet $(w, w^{'})_L$ has
the opposite $X$ charge and hypercharge to $(t, b)_L$
while the right-handed doublet has zero $X$ charge.
These acquire mass from a complex weak singlet Higgs scalar.
The electric charges of these {\it weird} quarks are $+1/3$ and $-2/3$;
they thus give rise to stable fractionally-charged color
singlets which may be problematic cosmologically. An alternative anomaly cancellation
is to add quark $SU(2)$ doublets, with $Y = + 1/6$,  $(t^{'},b^{'})_L (X= -1)$ + 
$(t^{'},b^{'})_R (X=0)$ together with
$SU(2)$ singlet $Y = -1$ charged leptons $l_L^{-} (X=1) + l_R^{-} (X=0)$ and $l_L^{-} (X= -1) + l_R^{-} (X=0)$. 

There is a three-dimensional parameter space for the model spanned by
$\tan\alpha$, $g_X$ and $\xi = \hat{M}_Z/\hat{M}_{Z^{'}}$.
We consider, for simplicity, only $\hat{M}_Z < \hat{M}_{Z^{'}}$
and will be able to constrain these parameters.  Using the analysis above we
have from the constraint on $R_b$,

\begin{equation}
0.008 \ge g_X \rm{tan}\alpha \ge 0.004~,\label{gxtana}
\end{equation}
as well as a weaker constraint from the asymmetry: $ g_X \tan\alpha < 0.07 $.
Turning this around using the $\delta R_b$ constraint,
gives a {\it prediction} for the asymmetry:

\begin{equation}
3 \times 10^{-4}  \geq  \delta A_{FB}^{(0,b)}  \geq   2 \times 10^{-4}~.
\end{equation}
This will be detectable if the experimental accuracy can be increased by
a factor of at least $3$ to $5$.
The quantity $\tan\alpha$ can be further restricted by perturbativity
and by custodial ${\rm SU}(2)$.
An upper limit $g_X(M_Z) < \sqrt{4\pi} = 3.54$, combined
with the $\delta R_b$ constraint dictates that

\begin{equation}
\tan\alpha  >  0.001~.\label{tanpert}
\end{equation}
The accuracy of custodial ${\rm SU}(2)$ symmetry
(the $\rho$ parameter) in the presence of multiple $Z$'s can be
expressed in terms of $\rho_i = M_W^2 /(\hat{M}^2_{Z_i}c_W^2)$~\cite{langluo}.
With just two $Z$'s we have the relationship
\begin{equation}
\tan^2\alpha = \frac{\bar{\rho}_1 - 1}{\xi^{-2}-\bar{\rho}_1}~,\label{rho}
\end{equation}
where $\bar{\rho}_i = \rho_i / \hat{\rho}$ with $\hat{\rho} = 1 + \rho_t$
which takes into account the top quark radiative corrections.
Rewriting Eq.~(\ref{Zwidth}) in terms of the Fermi constant $G_F$,
we find that all the decay rates are multiplied by a factor of
$\bar{\rho}_{eff} = \bar{\rho}_1\cos^2\alpha$ compared to the SM.
Using the the global fit allowing new physics in $R_b$ from
Ref.~\cite{pdglang} we have $\bar{\rho}_{eff} =
1.0002 \pm 0.0013 \pm 0.0018$ and Eq.~(\ref{rho}) gives, for $\alpha \ll 1, \xi
\ll 1$,

\begin{equation}
\tan \alpha < 0.045{\xi\over\sqrt{1 - 2\xi^2}}~.\label{rhobound}
\end{equation}
Since we have the lower bound on $\tan\alpha$ from Eq.~(\ref{tanpert}),
we deduce that $\xi > 0.028$ implying that $\hat{M}_{Z^{'}} < 3.3$ TeV.  It
is very interesting that the present model produces such an {\it upper} limit
on the new physics because it implies its testability in the
next generation of accelerators.

Because we have assigned $X$-charge asymmetrically to the three families,
there is inevitably a violation of GIM suppression\cite{GIM} of the
Flavor-Changing Neutral Currents (FCNC). In fact, study of FCNC sharpens the
definition of our model. When we assigned $X_L^{t,b} = 1$, there was an
inherent
ambiguity  of basis for the left-handed doublet $(t, b)_L$ because in general
a unitary transformation is needed to relate this doublet to the mass
eigenstates. The two most predictive limiting cases, out of an infinite range,
are where
(i) $t$ (ii) $b$ in $(t, b)_L$ is a mass eigenstate.
If $t$ is a mass eigenstate, then the empirical\cite{PDG} value
$\Delta m_B = (3.4 \pm 0.4) \times 10^{-13}$GeV imposes an upper limit on
the product ($g_X\xi$) too small, to
be consistent with the necessary increase $\delta R_b$.  On the other hand, if
$b$  is a mass eigenstate the $Z^{'}$-exchange
contribution to $\Delta m_B$ vanishes as do the (less constraining)
FCNC effects like $\Delta m_K$, $b \rightarrow s\gamma$, $b \rightarrow
s\bar{l}l$.

The model with $b$ a mass eigenstate can be made natural by imposing the
discrete symmetry $b_R \rightarrow - b_R, \phi' \rightarrow - \phi'$.  This
symmetry is spontaneously broken at the weak scale\footnote{ A second $\phi^{'}$ field 
is actually necessary to avoid an undesirable spontaneously-broken global $U(1)$.}
 but because it suffers from
a QCD anomaly there is no domain wall problem\cite{preskill}.  With the
discrete symmetry the Yukawa couplings of the neutral components of the Higgs
doublets are
\begin{equation}
{\cal L} = g_t \bar t_L t_R \phi^{(0)''} + g_b \bar b_L b_R \phi^{(0)'*} +
g_{ij}^{(u)} \bar u_{iL} u_{jR} \phi^{(0)*} + g_{ij}^{(d)} \bar d_{iL} d_{jR}
\phi^{(0)} + g_{i3}^{(u)} \bar u_{iL} t_R \phi^{(0)*} + h.c.,
\end{equation}
where $\alpha,\beta \in \{1,2,3\}$ and $\{i,j\} \in \{1,2\}$ (we neglect 
the exotic fermions Yukawa couplings to the ordinary ones).  The
weak eigenstate quark fields are related to primed mass eigenstate fields by
\[
u_L = U_L^\dagger u_L'~~ ~~d_L = T_L^\dagger d_L'\]
\begin{equation}
u_R = U^\dagger_R u'_R ~~ ~~  d_R = T_R^\dagger d_R'
\end{equation}
where $T_{33} = 1$ and $T_{3i} = T_{i3} = 0$.  The Kobayashi--Maskawa matrix
that occurs in the charged $W$ boson couplings, $(g_2/\sqrt{2}) \bar u'_{\alpha
L}  \gamma_\mu V_{\alpha\beta} d'_{\beta L} W^\mu$, is
\begin{equation}
V_{\alpha\gamma} = U_{L\alpha\beta} T_{L\beta\gamma}^\dagger
\end{equation}
implying that $V_{\alpha 3} = U_{L\alpha 3}$ and $V_{\alpha j} = U_{L\alpha i}
T_{Lij}$.  It follows that the flavor changing $Z'$ boson couplings are
\begin{equation}
{\cal L}_{FCNC} = g_X Z'_\mu (\bar u_{\alpha L}' \gamma^\mu V_{\alpha 3}
V^*_{\beta 3} u'_{\beta L})
\end{equation}
and that the flavor changing neutral Higgs boson couplings are
\begin{equation}
{\cal L}_{FCNC} = \left({m_t\over v^{\prime\prime}}\right)
\left(\phi^{(0)\prime\prime} - {v^{\prime\prime}\over v} \phi^{(0)*}\right)
\bar u_{L\alpha}' V_{\alpha 3} U^*_{R3\beta} u'_{R\beta}.
\end{equation}
The chief FCNC constraint now comes from the experimental bound
\cite{PDG} $\Delta m_D < 1.3 \times 10^{-13}$GeV.
The $Z^{'}$-exchange contribution gives 

\begin{equation}
\delta(\Delta m_D)
\simeq (g_X\xi)^2(7 \times 10^{-6}GeV) Re [V_{13}
V_{23}^*]^2(f_D/(0.22GeV))^2
\end{equation}
and hence requires
instead only a mild constraint $g_X\xi < 1$, easily consistent with $\delta
R_b$.  There is also a contribution to ($\Delta m_D$) from neutral Higgs
exchange
but the neutral Higgs masses can be chosen so that this is acceptably
small.
For example, the $\phi-$ and $\phi^{''}-$ exchange contribution to $D\bar{D}$
mixing is sufficiently suppressed (by third-family mixing) to allow Higgs
masses
$\simeq  250$GeV.

Fitting the hadronic width of $Z$ in our model gives rise to a {\it decrease}
in $\alpha_s(M_Z)$ and tends to
resolve discrepancies with low-energy determinations.

\vspace*{0.7cm}

{\bf III. Testability.}

Now let us consider the production of $Z^{'}$ in colliders. In $p\bar{p}
\rightarrow Z^{'}X$, the $Z^{'}$ is dominantly produced in association
with two $b$ quarks. The cross-section at $\sqrt{s} = 1.8$ TeV
falls off rapidly with $M_{Z^{'}}$: for example,
putting $g_X = g_Z$, it decreases from $16$ pb at $M_{Z^{'}} = 100$ GeV
to $1$ fb at $M_{Z^{'}} = 450$ GeV.
Against the $b\bar{b}$ background from QCD such a signal would be
difficult to observe at Fermilab. In particular, $Z^{'}$ production
leads to final states with four heavy-flavor jets and one
expects competition from QCD jet production to be severe.
At an $e^+e^-$ collider, sitting at the $Z^{'}$-pole, there is a
possibility for detecting the $Z^{'}$.
The coupling to $e^+e^-$ is
suppressed by tan$\alpha$ but still the pole can show up above background. 

\vspace*{0.9cm}

{\it To exhibit this effect in figures, the new data mentioned in Section I
are first accommodated by changing
the details of the model. The data are fitted within $\pm 1.3\sigma$
by assigning the quantum numbers:

\begin{equation}
X^{b,t}_{L,R} = 0, X^b_l = +1, X^t_R = -1.
\end{equation}
We set $g_Xtan\alpha = -0.018$.
Anomaly cancellation can be accomplished by adding SU(2) singlet 
quarks $b^{'},t^{'}$ with charges $X_L^{b^{'}} = -1, X^{b^{'}}_R = 0,
X_L^{t^{'}} = +1, X_R^{t^{'}} = 0$.
The SU(2)
doublet scalars $\phi, \phi^{'}, \phi^{''}$ have $X = 0, +1, -1$ respectively
as in the text.
There follow figures which show (Fig.~1) 
$\sigma(e^+e^- \rightarrow b\bar{b})$, (Fig.~2) $A^b_{FB}$ and
(Fig.~3) $A^b_{LR}$ for $M(Z^{'}) = 250 {\it\ and\ } 150$ GeV. 
The other relevant model parameters are shown in the plots.
}

\begin{figure}[htb]
\psfull
\begin{center}
\mbox{\epsfig{file=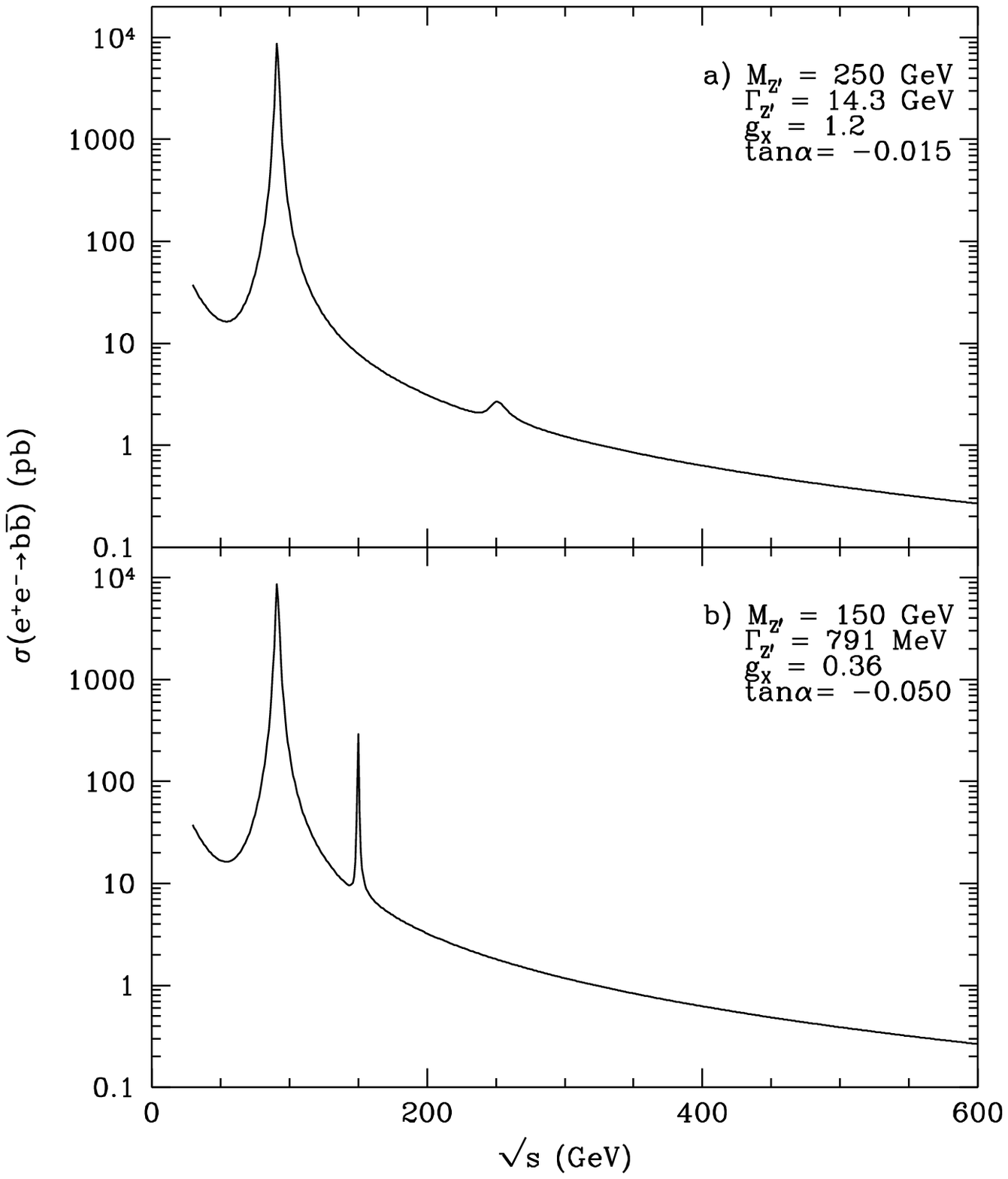}}
\end{center}
\fcaption{Cross section for $e^+e^- \rightarrow \bar{b}b$
for $Z^{'}$ masses a) 250 GeV and b) 150 GeV and model parameters
shown in the plots.}
\end{figure}

\begin{figure}[htb]
\psfull
\begin{center}
\mbox{\epsfig{file=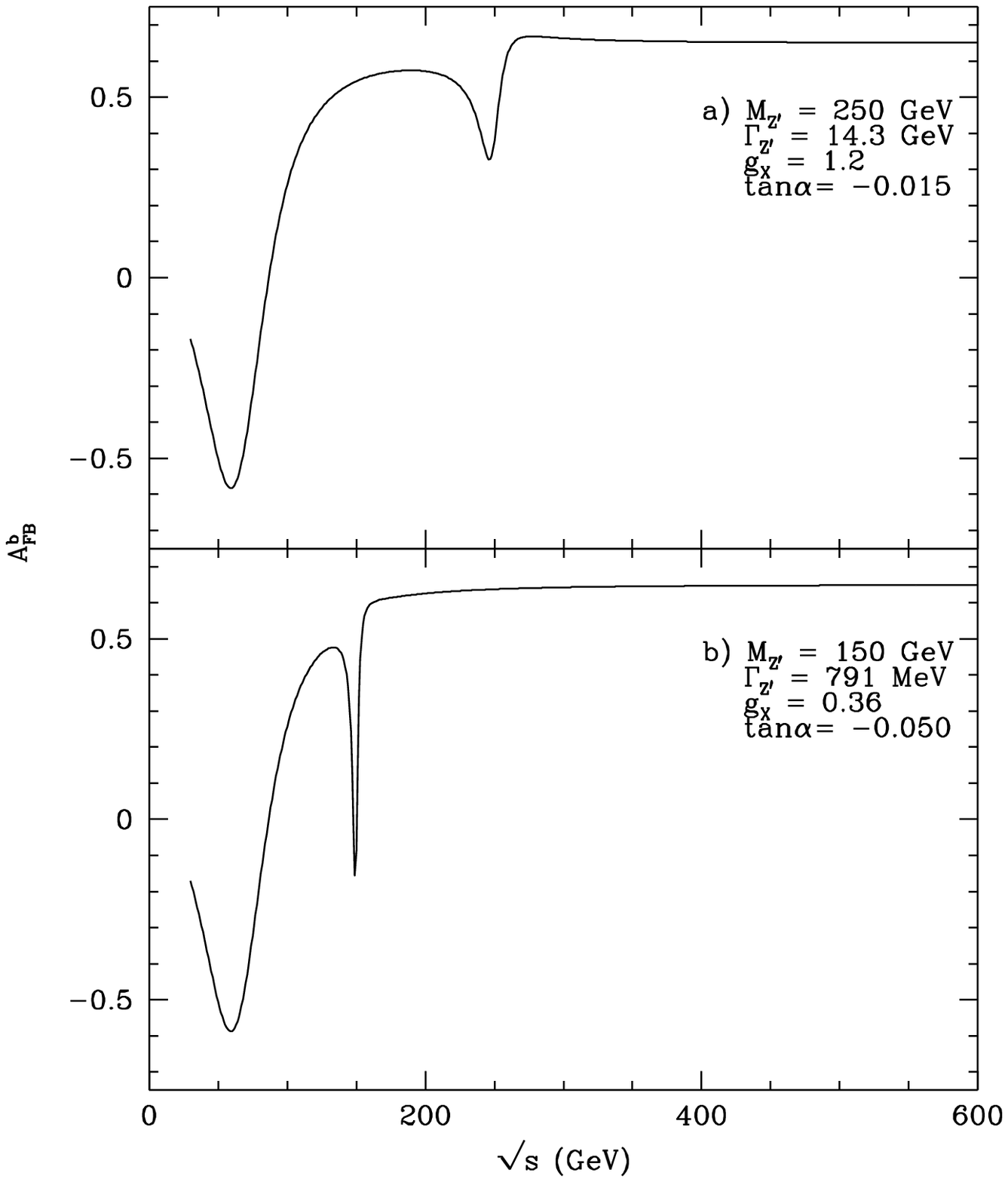}}
\end{center}
\fcaption{Forward-backward asymmetry in 
$e^+e^- \rightarrow \bar{b}b$
for $Z^{'}$ masses a) 250 GeV and b) 150 GeV and model parameters
shown in the plots.}
\end{figure}

\begin{figure}[htb]
\psfull
\begin{center}
\mbox{\epsfig{file=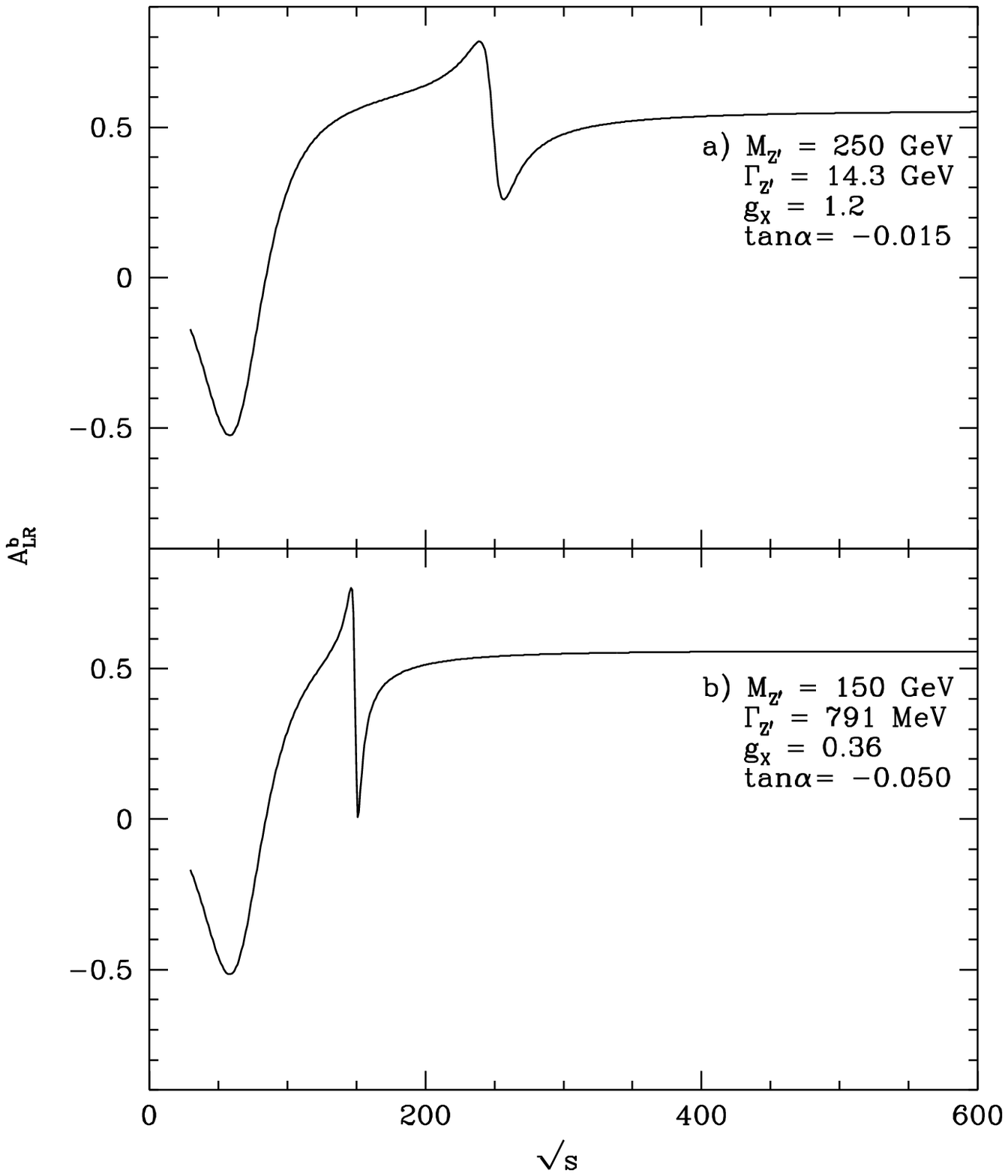}}
\end{center}
\fcaption{Left-right asymmetry in $e^+e^- \rightarrow \bar{b}b$
for $Z^{'}$ masses a) 250 GeV and b) 150 GeV and model parameters
shown in the plots.}
\end{figure}

\clearpage

{\bf IV. Summary.}

In summary, we have constructed a model which can account for the
measured value of $R_b$ and $A^b$. It introduces a 
$Z^{'}$ coupled almost entirely to the third family and to exotic fermions.
The model's interest is that $Z^{'}$ couples with
sizeable strength to $b$ and $t$ quarks
and can naturally avoid FCNC without a GIM mechanism.
$D\bar D^0$ mixing may be near its experimental value.
This $Z^{'}$ is particularly elusive
because it is so difficult to detect at colliders ---
with the possible exception of $e^+e^- \rightarrow \bar{b}b$
at the $Z^{'}$ pole. For example, if the $Z^{'}$ lies 
below 192GeV, the maximum energy envisaged for LEP2, the 
cross-section $e^+e^- \rightarrow b\bar{b}$ and asymmetry
$A^b_{FB}$ have readily detectable changes at the $Z^{'}$ pole.

\vspace*{0.9cm}

{\bf Acknowledgements.}

\vspace*{0.5cm}

I would like to thank M. Wise and B. Wright for their collaboration and 
M. Pohl for helpful discussions about the detectability of $Z^{'}$.
This work was supported in part by the U.S.
Department of Energy  under Grant No. DE-FG05-85ER-40219, Task B.

\vspace*{0.9cm}

{\bf References.}


\vspace*{0.5cm}

\begin{thebibliography}{99}

\bibitem{pdglang}
J. Erler and P.~Langacker, Phys. Rev. {\bf D52}, 441 (1995)
and articles by these authors in Ref.~\cite{PDG}.

\bibitem{PDG}
Review of Particle Properties, Particle Data Group, Phys. Rev. {\bf D50},
1173 (1994) and 1995 off-year partial update available at the URL:
{\it http://pdg.lbl.gov/}.

\bibitem{BBM}
P. Bamert, McGill University preprint MCGILL-95-64, Dec. 1995,
{\it hep-ph/9512445}; \\P. Bamert, C. P. Burgess and I. Maksymyk,
Phys. Lett. {\bf B356}, 282 (1995).

\bibitem{holdom}
B.~Holdom, Phys. Lett. {\bf B339}, 114 (1994).

\bibitem{CR}
F. Caravaglios and G. G. Ross, Phys. Lett. {\bf B346}, 159 (1995).

\bibitem{CA}
P. Chiappetta, J. Layssac, F. M. Renard and C. Verzegnassi,
Centre de Physique Th\'eorique preprint CPT-96/P.3304, Jan. 1996,
{\it hep-ph/9601306};\\
G. Altarelli, N. Di Bartolomeo, F. Feruglio, R. Gatto and M. Mangano,
CERN preprint CERN-TH/96-20, Jan. 1996, {\it hep-ph/9601324};\\
K. S. Babu, C. Kolda and J. March-Russell,  Phys. Rev. {\bf D54}, 4635 (1996).

\bibitem{ma}
E. Ma, Phys. Rev. {\bf D53}, 2276 (1996).

\bibitem{LN}
J. T. Liu and D. Ng, Phys. Lett {\bf B342}, 262 (1995).

\bibitem{BBCLN}
P. Bamert, C. P. Burgess, J. M. Cline, D. London and E. Nardi,
Phys. Rev. {\bf D54}, 4275 (1996). 

\bibitem{WK}
J. D. Wells and G. Kane, Phys. Rev. Lett. {\bf 76}, 869 (1996).

\bibitem{ELN}
J. Ellis, J. L. Lopez and D. V. Nanopoulos, CERN preprint CERN-TH/95-314,
Dec. 1995, {\it hep-ph/9512288}.

\bibitem{FPT}
L. Clavelli, Mod. Phys. Lett. {\bf A10}, 949 (1995);\\
J. L. Feng, N. Polonsky and S. Thomas, SLAC preprint SLAC-PUB-95-7050,
Nov. 1995, {\it hep-ph/9511324}.

\bibitem{TH}
E. Ma and D. Ng, Phys. Rev. {\bf D53}, 255 (1996).

\bibitem{barger}
K. Agashe, M. Graesser, I. Hinchliffe and M. Suzuki, LBL-38569 (1996),
{\it hep-ph/9604266}.\\
V. Barger, K. Cheung and P. Langacker, MADPH-96-936/ UPR-0696T/ UTEXAS-HEP-96-2/ DOE-ER-40757-078 (1996), {\it hep-ph/9604298}.

\bibitem{wise}
P. H. Frampton, M. B. Wise and B. Wright, Phys. Rev. {\bf D}(Nov. 1, 1996 issue),
{\it hep-ph/9604260.}

\bibitem{langluo}
P. Langacker and M. Luo, Phys. Rev. {\bf D45}, 278 (1992).

\bibitem{GIM}
S.~L.~Glashow, J.~Iliopoulos and L.~Maiani, Phys. Rev. {\bf D2}, 1285 (1970).

\bibitem{preskill}
J. Preskill, S. Trivedi, F. Wilczek, and M.B. Wise, Nucl. Phys. {\bf B363}, 207
(1991).


\end{thebibliography}
\end{document}